\documentstyle[12pt]{article}
\begin{document}

\begin{center}
{\bf{q-Fermionic Numbers and Their Roles in Some Physical Problems}}

\vspace{0.5cm}

R.Parthasarathy{\footnote{e-mail address: sarathy@imsc.res.in}}   \\
The Institute of Mathematical Sciences \\
C.I.T. Campus, Tharamani Post \\
Chennai, 600 113, India. \\
\end{center} 

\vspace{0.5cm}

{\noindent{\it{Abstract}}}

\vspace{0.5cm}

The q-fermion numbers emerging from the q-fermion oscillator algebra are used to
reproduce the q-fermionic Stirling and Bell numbers. New recurrence
relations for the expansion coefficients in the 'anti-normal ordering' of the
q-fermion operators are derived. The roles of the q-fermion numbers in q-stochastic
point processes and the Bargmann space representation for q-fermion operators are
explored.  

\vspace{0.5cm}

\newpage  
q-deformed Stirling numbers were introduced by Carlitz, Gould and Milne [1]. 
Such numbers are encountered in [2] in the 
'normal ordering' of q-deformed boson oscillator creation and annihilation operators [3]. Recently, Schork
[4] considered generalized q-Stirling numbers and obtained useful properties. He further introduced
unsigned q-deformed Lah numbers. Subsequently, he [5] considered q-fermionic Stirling numbers which are
encountered in the 'normal ordering' of q-fermionic oscillator creation and annihilation operators
introduced by the author and Viswanathan [6]. 
Katriel [7] has derived q-Dobinski formula for q-bosonic Bell number by using q-boson coherent states. In
contrast to this semi-classical derivation, a probabilistic derivation of the q-Dobinsky formula for
q-bosonic Bell number has been obtained by the author and Sridhar [8] by considering q-stochastic point
processes.

\vspace{0.5cm}
 
The Stirling numbers are crucial in many combinatorial problems [9]. Milne [1] in his studies on
generalized restricted growth functions obtained q-Stirling numbers (bosonic), q-Dobinski formula and
q-Charlier polynomials. It is very intriguing that these numbers occurring in combinatorial problems,
arise naturally in undeformed and q-deformed harmonic oscillator algebra of the creation and annihilation
operators.

\vspace{0.5cm}

In this paper we consider q-fermionic numbers in more detail and obtain q-fermionic Stirling 
and Bell numbers {\it{explicitly}}. 
 Further we consider the 'anti-normal ordering' of the q-deformed bosonic and
fermionic operators and obtain new recurrence relations. Using their Fock space
states, we give a meaning to these numbers as expressing powers of q-bosonic and q-fermionic numbers in
terms of 'raising factorials'. This is the counterpart of the q-Stirling numbers of the second kind in
expressing the powers of q-numbers in terms of 'falling factorials'.         
 
In the subsequent part of this paper, we attempt to give a probabilistic interpretation of the q-fermionic
Stirling numbers of the second kind by considering q-stochastic point processes. This introduces q-product
densities, a generalization of the concept of product densities introduced in 1950 by Ramakrishnan [10] in his
study of cosmic ray cascades.  
Then, we give a
Bargmann space representation of the q-fermion operators as multiplication by and q-differentiation with respect
to quasi-Grassmann variable leading to  
 differential equations involving
q-differentiation on spaces of entire functions of quasi-Grassmann variable.   

\vspace{0.5cm}

{\noindent{\bf{2.q-fermionic numbers}}}

\vspace{0.5cm}

Macfarlane [3] and Biedenharn [3] introduced q-boson oscillator algebra as
\begin{eqnarray}
aa^{\dagger}-\sqrt{q}a^{\dagger}a=q^{-N/2} &;& [N,a]=-a\ \ ,\ \ [N,a^{\dagger}]=a^{\dagger},
\end{eqnarray}
where $q>0$. By making a transformation
\begin{eqnarray}
A=q^{N/4}a &;& A^{\dagger}=a^{\dagger}q^{N/4},
\end{eqnarray}
one obtains
\begin{eqnarray}
AA^{\dagger}-qA^{\dagger}A&=& 1;\ \ q>0,
\end{eqnarray}
and the associated q-bosonic number
\begin{eqnarray}
{[n]_b}&=& \frac{1-q^n}{1-q}.
\end{eqnarray}
In the above {\it{$q$ is strictly positive}}. The author and Viswanathan [6], proposed
a non-trivial q-fermion oscillator algebra as
\begin{eqnarray}
ff^{\dagger}+\sqrt{q}f^{\dagger}f=q^{-N/2} &;& [N,f]=-f, [N,f^{\dagger}]=f^{\dagger} ,  f^2\neq \ 0\ , 
\ (f^{\dagger})^2\neq 0,  
\end{eqnarray}
where $q>0$. By making  a transformation 
\begin{eqnarray}
F=q^{N/4}f &;& F^{\dagger}=f^{\dagger}q^{N/4},
\end{eqnarray}
one obtains
\begin{eqnarray}
FF^{\dagger}+qF^{\dagger}F=1 &;& F^2\neq 0,\ (F^{\dagger})^2\neq 0,\ q>0,
\end{eqnarray}
and the associated q-fermionic number 
\begin{eqnarray}
{[n]_f}&=& \frac{1-(-1)^nq^n}{1+q}.
\end{eqnarray}
In (5) to (8), $q$ is strictly positive. {\it{Thus, it is not correct to replace $q$ by $-q$ in the q-boson
algebra to obtain q-fermion algebra (7), as $q$ in (1) to (4) is strictly positive.}} 
Nevertheless, for
mathematical expressions such replacement may be carried out. In what follows, we shall denote (8) by q-fermion
number. This definition is of fundamental importance and is quite different from $[n]_b$. In the limit
$q\rightarrow 1$, $[n]_b \rightarrow n$ while $[n]_f \rightarrow \frac{1}{2}(1-(-1)^n)$ taking values $0$ and $1$
for $n$ even and odd. 
The properties of the q-fermion numbers associated with (5) (which are different from (8))
have been studied by Narayana Swamy [11]. For $q<1$, the q-fermion numbers (8) {\it{never}} go beyond 1 for any
value of $n$ and for $n\rightarrow \infty $, it asymptotically approches $0.5$. 
On the other hand the q-boson
numbers (4), become the usual numbers when $q=1$, and for $q<1$, 
as $n\rightarrow \infty $ we have, $ [n]_b$ asymptotically
goes to $1/(1-q)$. For $q>1$, we have for q-fermion numbers, $[0]_f=0\ ,\ [1]_f=1$ and $[n]_f>0$ for $n$ odd, $<0$
for $n$ even. 

The q-fermions described by (5) or (7) with the q-numbers (8) are
different from the k-fermions introduced by Daoud, Hassouni and Kibler [12]. The
k-fermion algebra is a non-Hermitian realization of the q-deformed Heisenberg bosonic
algebra with $q$ being a root of unity and satisfy $f^k_{\pm}=0$ and $f_{+}\neq {f_{-}}^{\dagger}$, except for $k=2$, for which they become ordinary fermions. On the other hand,
q-fermions admit $q$ real or complex and only when $q=1$, they become ordinary
fermions.    

\vspace{0.5cm}

{\noindent{\bf{3.Normal Ordering of q-fermion operators}}}

\vspace{0.5cm}

We wish to evaluate $(F^{\dagger}F)^r$ using (7). It is straightforward to expand,
\begin{eqnarray}
(F^{\dagger}F)^r&=& \sum_{s=1}^r {\cal{F}}^r_s (F^{\dagger})^sF^s,
\end{eqnarray}
and find a recurrence relation for ${\cal{F}}^r_s$. From (9), it follows
\begin{eqnarray}
(F^{\dagger}F)^{r+1}
     &=& \sum_{s=1}^r {\cal{F}}^r_s (F^{\dagger})^sF^s F^{\dagger}F.
\end{eqnarray}
From (7), we have
\begin{eqnarray}
F^sF^{\dagger}&=&[s]_f F^{s-1} + (-1)^s q^s F^{\dagger}F^s.
\end{eqnarray}
Using (11) in (10) and noting from (9), ${\cal{F}}^r_0=0\ ,\ {\cal{F}}^r_{r+1}=0$, we find 
\begin{eqnarray}
{\cal{F}}^{r+1}_s&=&(-1)^{s-1}q^{s-1} {\cal{F}}^r_{s-1}+[s]_f {\cal{F}}^r_s,
\end{eqnarray}
the desired recurrence relation for ${\cal{F}}^r_s$ coefficients in (9). 
 The recurrence relation (12) with  
${\cal{F}}^1_1=1$ is the same as that of the q-fermionic Stirling numbers of the second kind
[5]. 

\vspace{0.5cm}

The q-fermionic Bell number introduced in [5] is
\begin{eqnarray}
{\cal{B}}^{(f)}_r&=&\sum_{s=1}^r {\cal{F}}^r_s.
\end{eqnarray}
An attempt along the lines of [7] for q-fermionic Bell number runs into difficulty. q-fermion coherent
states have been constructed in [13] using 'quasi Grassmann' variables $\psi$. It is to be noted that the
replacement of $q$ by $-q$ in the q-boson coherent states {\it{will not}} give q-fermion coherent states. 
These two
coherent states are structurally very different.  
As ${\psi}^{\dagger}\psi+\psi {\psi}^{\dagger}=0; {\psi}^2\neq 0, ({\psi}^{\dagger})^2\neq 0$, it is not possible
to use the analogue of $|z|=1$ here. 
So, we take the matrix elements of (9) between q-fermion Fock space states [13] $|n>$ with $n>r$ and use
$F|n>=\sqrt{[n]_f}|n-1>\ ;\ F^{\dagger}|n>=\sqrt{[n+1]_f}|n+1>$ to arrive at  
\begin{eqnarray}
{[n]_f}^r &=& \sum_{s=1}^r {\cal{F}}^r_s \frac{[n]_f!}{[n-s]_f!},
\end{eqnarray}
which can be verified explicitly using (8) and (12). Multiplying (14) by ${\lambda}^n$ and summing $n$ from $1$ to
$\infty$ and then setting $\lambda =1$, we obtain
\begin{eqnarray}
{\cal{B}}^{(f)}_r &=& (e_q^{(f)}(1))^{-1}\sum_{n=1}^{\infty} \frac{[n]_f^r}{[n]_f!},
\end{eqnarray}
where 
\begin{eqnarray}
e^{(f)}_q(x)&=&\sum_{n=0}^{\infty} \frac{x^n}{[n]_f!}. \nonumber 
\end{eqnarray}        
(15) is the q-fermionic Dobinski formula. 
 
\vspace{0.5cm}

Some of the q-fermionic Bell numbers are:
\begin{eqnarray}
{\cal{B}}^{(f)}_1&=&1, \nonumber \\
{\cal{B}}^{(f)}_2&=&1-q, \nonumber \\
{\cal{B}}^{(f)}_3&=&1-q-q[2]_f-q^3, \nonumber \\
{\cal{B}}^{(f)}_4&=&1-q-q[2]_f-q[2]_f^2-q^3-q^3[2]_f-q^3[3]_f+q^6, \nonumber \\
{\cal{B}}^{(f)}_5&=&1-q+(-q-q[2]_f-q[2]_f^2)([2]_f+q^2) \nonumber \\
        &+&(-q^3-q^3[2]_f-q^3[3]_f)([3]_f-q^3)+q^6[4]_f+q^{10}.
\end{eqnarray}
In the limit $q=1$, we have
\begin{eqnarray}
{\cal{B}}^{(f)}_1=1 &;& {\cal{B}}^{(f)}_2=0, \nonumber \\
{\cal{B}}^{(f)}_r&=& (-1)^r\ \ \ \ \ \ if\ r=0\ (mod\ 3) \nonumber \\
                 & & (-1)^{r+1}\ \ \ if\ r=1\ (mod\ 3) \nonumber \\
                 & & 0 \ \ \ \ \ \ \ \ \ \ \  if \ r=2\ (mod\ 3).
\end{eqnarray}
The results (17) have been obtained by Wagner [14] in his study of generating functions for some well known
statistics on the family of partitions of a finite set. 

\vspace{0.5cm}

The q-fermionic Stirling numbers of the {\it{first}} kind are introduced by expressing the inverse of (9), as  
\begin{eqnarray}
(F^{\dagger})^rF^r&=&\sum_{s=1}^r {\cal{S}}^r_s (F^{\dagger}F)^s. \nonumber 
\end{eqnarray}    
The matrix elements of the above between q-fermion Fock space states $|n>, (n>r)$ give  
\begin{eqnarray}
\frac{[n]_f!}{[n-r]_f!}&=& \sum_{s=1}^r {\cal{S}}^r_s [n]_f^s. 
\end{eqnarray}
Using $[n]_f-[r]_f=(-1)^rq^r[n-r]_f$, we have the recurrence relation
\begin{eqnarray}
{\cal{S}}^{r+1}_s&=&(-1)^rq^{-r}{\cal{S}}^r_{s-1}-[r]_f(-1)^rq^{-r}{\cal{S}}^r_s.
\end{eqnarray}
Similarly, the q-fermionic unsigned Lah numbers introduced as
\begin{eqnarray}
\frac{[r+n-1]_f!}{[r-1]_f!}&=&\sum_{s=0}^n {\cal{L}}^n_s \frac{[r]_f!}{[n-s]_f!},
\end{eqnarray}
have the recurrence relation
\begin{eqnarray}
{\cal{L}}^{n+1}_s&=&(-1)^{n+s-1}q^{n+s-1}{\cal{L}}^n_{s-1}+[s+n]_f{\cal{L}}^n_s.
\end{eqnarray}
These mathematical results reveal the feature of obtaining them from their q-bosonic counterparts by replacing $q$
by $-q$. The unsigned q-Lah numbers are met in the normal ordering of $((A^{\dagger})^rA^s)^n$ for $r=2$ and $s=1$
[15] and their
q-analogues have been obtained in [4,5]. 

\vspace{0.5cm}

{\noindent{\bf{4.Anti-Normal Ordering of q-fermion operators}}}

\vspace{0.5cm}

In this section, we seek an expansion for $(AA^{\dagger})^r$. First, we consider q-bosonic operators in (3). It is
straightforward to expand  
\begin{eqnarray}
(AA^{\dagger})^r&=&\sum_{s=1}^r {\cal{A}}^r_s A^s (A^{\dagger})^s,
\end{eqnarray}
and use 
\begin{eqnarray}
(A^{\dagger})^sA&=&\frac{1}{q^s}A(A^{\dagger})^s-\frac{1}{q^s}[s](A^{\dagger})^{s-1}, \nonumber 
\end{eqnarray}
to obtain a recurrence relation for ${\cal{A}}^r_s$ with ${\cal{A}}^1_1=1, {\cal{A}}^r_0=0, {\cal{A}}^r_{r+1}=0$
as 
\begin{eqnarray}
{\cal{A}}^{r+1}_s&=&q^{-(s-1)}{\cal{A}}^r_{s-1}-[s]q^{-s}{\cal{A}}^r_s.
\end{eqnarray}            

\vspace{0.5cm}

A similar relation for q-fermionic operators anti-normal orderering, namely
\begin{eqnarray}
(FF^{\dagger})^r&=&\sum_{s=1}^r {\cal{B}}^r_s F^s (F^{\dagger})^s,
\end{eqnarray}
can be obtained from (7). We have from (7) 
\begin{eqnarray}
(F^{\dagger})^sF&=&(-1)^sq^{-s}F(F^{\dagger})^s-(-1)^sq^{-s}[s]_f(F^{\dagger})^{s-1}. \nonumber 
\end{eqnarray}
Using this and (24), we find $( {\cal{B}}^1_1=1; {\cal{B}}^r_0=0; {\cal{B}}^r_{r+1}=0)$,
\begin{eqnarray}
{\cal{B}}^{r+1}_s&=&(-1)^{s-1}q^{-(s-1)}{\cal{B}}^r_{s-1}-(-1)^sq^{-s}[s]_f{\cal{B}}^r_s.
\end{eqnarray}
These recurrence relations (23) and (25) are {\it{different}} from those of q-Stirling or Lah numbers. 

\vspace{0.5cm}

In order to obtain a relationship between ${\cal{A}}^r_s\ ({\cal{B}}^r_s)$ and $[n]_b \ ([n]_f)$, we 
 use  the q-boson Fock space states for (22) and q-fermion Fock space for (24).
Then using the standard action of the creation and annihilation q-operators on the Fock
space states, we find  
\begin{eqnarray}
{[n+1]_b^r}&=&\sum_{s=1}^r {\cal{A}}^r_s \frac{[n+s]_b!}{[n]_b!}, \nonumber \\ 
{[n+1]_f^r} &=& \sum_{s=1}^r {\cal{B}}^r_s \frac{[n+s]_f!}{[n]_f!}.
\end{eqnarray}
It is interesting to observe that {\it{while the normal ordering of operators yielded expressions for $[n]_b^r \
([n]_f^r)$ in terms of q-stirling numbers of the second kind as 'falling factorials' (namely (14) and its
q-bosonic analogue), the anti-normal ordering yields expressions for $[n]_b^r ([n]_f^r)$ in terms of
${\cal{A}}^r_s,\ {\cal{B}}^r_s$ as 'raising factorials'}}.        

\vspace{0.5cm}

{\noindent{\bf{5.q-fermionic Stirling number of second kind - a probabilistic view}}}

\vspace{0.5cm}

In this section we extend the theory of product densities of Ramakrishnan [10] to a q-extension of the stochastic
variable $[n(E)]_f$ which depend on a continuous parameter $E$ taken to be ordinary variable, The statistical
properties of these q-fermionic stochastic variable taking values $[n(E)]_f$ will be governed by q-stochastic
point processes. Recall that $[n]_f$ for $q<1$ never exceeds unity. So the feature assumed in [10] namely atmost
one particle occurs in the interval $dE$ is maintained. Now we {\it{define}} the q-number of particles in the
range $E$ and $E+dE$ to be $[n(E+dE)-n(E)]_f$ which is just $[dn(E)]_f$. Following [10], we take the probability
that there occurs $[1]_f=1$ particle in the interval $dE$ is proportional to $dE$ and that for the occurrence of
$[n]_f$ particles in $dE$ is proportional to $(dE)^n$. The average number of particles in the interval $dE$,
denoted by ${\cal{E}}([dn(E)]_f)$, is represented by a q-function $f^{(q)}_1(E)$ such that 
\begin{eqnarray}
{\cal{E}}([dn(E)]_f)&=&f^{(q)}_1(E)\ dE. 
\end{eqnarray}
Denoting the probability that $[n]_f$ particles occuring in the interval $dE$ by $P_q([n]_f)$, we have 
\begin{eqnarray}
P_q([1]_f)&\equiv &P_q(1)\ =\ f^{(q)}_1(E)dE+{\cal{O}}((dE)^2), \nonumber \\
P_q([0]_f)&\equiv &P_q(0)\ =\ 1-f^{(q)}_1(E)dE-{\cal{O}}((dE)^2), \nonumber \\
P_q([n]_f)&=&{\cal{O}}((dE)^n)\ ;\ n>1.
\end{eqnarray}     

\vspace{0.5cm}

The average of the $r^{th}$ moment of $[n]_f$ is then 
\begin{eqnarray}
{\cal{E}}([n]^r_f)&=&\sum_n [n]^r_f P_q([n]), \nonumber \\
                  &=&f^{(q)}_1(E)dE\ =\ {\cal{E}}([dn(E)]),
\end{eqnarray}
where the second step follows from (27) and (28). Thus all the moments are equal to the probability that the
q-stochastic variable assumes the value $[1]_f=1$. This feature of [10] is maintained here. $f^{(q)}_1(E)$ is the
q-product density of degree $1$.

\vspace{0.5cm}

Now we consider the distribution of $[n]_f$ particles in the $E$-axis, that is, in the intervals $dE_1, dE_2,
\cdots dE_n$, with $[1]_f (=1)$ particle in each interval. For the first interval $dE_1$, this can be done in
$^{[n]_f}C_{[1]_f}\ =\ [n]_f!/{[n-1]_f!}\ =\ [n]_f$ {\it{number of ways}}, where $^{[n]_f}C_{[1]_f}$ is the
q-binomial coefficient. Thus, from the average number of particles in (27), we have
\begin{eqnarray}
f^{(q)}_1(E_1)dE_1&=&[n]_f f^{(q)0}_1(E_1)dE_1,
\end{eqnarray}
with 
\begin{eqnarray}
\int_{whole\ range} f^{(q)0}_1(E)dE&=&1.
\end{eqnarray}
If we now use the reamaining particles as $([n]_f-1)$, then the number of ways of putting $[1]_f$ particle in $dE_2$
will be $[n]_f-1$. Then the joint probability of putting $[1]_f$ particle {\it{each}} in $dE_1$ and $dE_2$ will be
proportional to $[n]_f([n]_f-1)$, if we were to use the same product density in [10]. But, in this way, we will
not be exhausting the total number $[n]_f$ of particles. So, in dealing with q-numbers, {\it{it is necessary to
introduce q-product densities}}, such that the joint probability of putting $[1]_f$ particle {\it{each}} in $dE_1$
and $dE_2$ will be taken to be proportional to $[n]_f[n-1]_f$. Then the average number in this case will be 
\begin{eqnarray}
{\cal{E}}([dn(E_1)]_f[dn(E_2)]_f)&\equiv & f^{(q)}_2(E_1,E_2)dE_1\ dE_2, \nonumber \\
                   &=& [n]_f[n-1]_f f^{(q)0}_1(E_1)f^{(q)0}_1(E_2)\ dE_1\ dE_2.
\end{eqnarray}
 
\vspace{0.5cm}

Proceeding further, the joint probability of putting $[1]_f$ particle each in $dE_1, dE_2, \cdots \ dE_n$ will be
proprtional to $[n]_f!$, thereby exhausting the total number $[n]_f$ particles. This gives
\begin{eqnarray}
f^{(q)}_m(E_1,\cdots E_m)dE_1\cdots dE_m&=&\frac{[n]_f!}{[n-m]_f!} f^{(q)0}_1(E_1)\cdots f^{(q)0}_1(E_m)dE_1\cdots
dE_m, \nonumber \\
f^{(q)}_n(E_1\cdots E_n)dE_1\cdots dE_n&=&[n]_f!f^{(q)0}_1(E_1)\cdots f^{(q)0}_1(E_n)dE_1\cdots dE_n. 
\end{eqnarray}
In (32) and (33), the intervals do not overlap. When the intervals overlap, a degeneracy occurs [10] and then for
a finite interval $\bigtriangleup E=E_u-E_t$
\begin{eqnarray}
\int_{E_t}^{E_u}\ \int_{E_t}^{E_u}{\cal{E}}([dn(E_1)]_f[dn(E_2)]_f)&=&\int_{E_t}^{E_u}f^{(q)}_1(E)dE \nonumber
\\
 &+& \int_{E_t}^{E_u}\ \int_{E_t}^{E_u} f^{(q)}_2(E_1,E_2)dE_1\ dE_2.
\end{eqnarray}
The $r^{th}$ moment of the q-number of particles in the finite range $\bigtriangleup E$, 
namely ${\cal{E}}([n]^r_{f\
\bigtriangleup E})$ can be represented, after taking the degeneracy into account, by
\begin{eqnarray}
{\cal{E}}([n]^r_{f\ \bigtriangleup E})&=&\sum_{s=1}^{r} {\cal{C}}^r_s \int_{E_t}^{E_u}\cdots \int_{E_t}^{E_u}
f^{(q)}_s(E_1,\cdots E_s)dE_1\cdots dE_s,
\end{eqnarray}
where the coefficients ${\cal{C}}^r_s$ are functions of $r$ and $s$ alone. For $[n]_f$ fixed, integrating over the
whole range and using (31) and (33), we obtain
\begin{eqnarray}
[n]^r_f &=& \sum_{s=1}^r {\cal{C}}^r_s \frac{[n]_f!}{[n-s]_f!},
\end{eqnarray}
which is same as (14) upon identifying ${\cal{C}}^r_s$ with ${\cal{F}}^r_s$. This derivation of (36) gives the
role of the q-fermionic Strirling number of the second kind as taking into account the degenarcaies in the joint
probabilities of the distribution of q-fermionic number $[n]_f$ as a stochastic variable when the intervals
overlap.  
    
\vspace{0.5cm}

{\noindent{\bf{6.Bargmann Space Representation for q-fermion operators}}}

\vspace{0.5cm}

 A Bargmann space realization for q-bosons has been developed by
Bracken, MacAnally, Zhang and Gould [16] and that for q-fermions has been developed by the author [17].
As the q-fermion coherent states involve quasi-Grassmann variable $\psi$, the space consists of monomials of
$\psi$. In this section, we briefly recall the main results to illustrate two points. First, a naive replacement
of $q$ by $-q$ is not sufficient and is incorrect. Second,
we would like to represent (9) and (24) in terms of 'differential operators'. The q-fermion coherent state [13]
is given by
\begin{eqnarray}
|\psi>&=& {\Big( e_q^{{\psi}^{\dagger}\psi}\Big)}^{-\frac{1}{2}} e_q^{-\psi F^{\dagger}}|0>, \nonumber \\
      &=& {\Big( e_q^{{\psi}^{\dagger}\psi}\Big)}^{-\frac{1}{2}}\sum_{n=0}^{\infty}(-1)^n\Bigl\{ \frac
{{\psi}^{2n}}{\sqrt{[2n]_f!}}|2n> \nonumber \\ 
 &-&\frac{{\psi}^{2n+1}}{\sqrt{[2n+1]_f!}}|2n+1>\Bigr\},
\end{eqnarray}
and it can be verified $F|\psi>=\psi |\psi>$. We will first map a vector in the Hilbert space $|\phi>$ to a 
function $\phi(\psi)$ by
\begin{eqnarray}
\phi(\psi) &\equiv & <{\psi}^{\dagger}|\phi>, \nonumber \\
           &=&{\Big( e_q^{{\psi}^{\dagger}\psi}\Big)}^{-\frac{1}{2}}\sum_{n=0}^{\infty}(-1)^n\Bigl\{
\frac{{\psi}^{2n}}{\sqrt{[2n]_f!}}<2n|\phi> \nonumber \\  
 &+&\frac{{\psi}^{2n+1}}{\sqrt{[2n+1]_f!}}<2n+1|\phi>\Bigr\},
\end{eqnarray}
where we have made use of the anti-commuting property of $\psi$ with $F$ and $F^{\dagger}$. In [17] it has been
shown that $\phi(\psi)$ is an entire function. 

\vspace{0.5cm}

Now consider the matrix element $<{\psi}^{\dagger}|F^{\dagger}|\phi>$. Using $F|\psi>=\psi |\psi>$, we have
\begin{eqnarray}
<{\psi}^{\dagger}|F^{\dagger}|\phi>&=& \psi \phi(\psi),
\end{eqnarray}
and so {\it{in the space of $\phi(\psi)$, $F^{\dagger}$ is represented by multiplication by $\psi$}}. Consider now
the expression $\psi <{\psi}^{\dagger}|F|\phi>$ which can rewritten as 
\begin{eqnarray}
\psi <{\psi}^{\dagger}|F|\phi>&=& <{\psi}^{\dagger}|F^{\dagger}F|\phi>.
\end{eqnarray}
Writing $F^{\dagger}F$ as $[N]_f$ (which is possible in view of the expression for $|\psi>$ in terms of expansion
of q-fermion Fock space states), we have 
\begin{eqnarray}
\psi <{\psi}^{\dagger}|F|\phi>&=& <{\psi}^{\dagger}|[N]_f|\phi>, \nonumber \\
                           &=& <{\psi}^{\dagger}|\frac{1-(-1)^Nq^N}{1+q}|\phi>,
\end{eqnarray}
where in the last step we used (8). Using (38), we have $<{\psi}^{\dagger}|q^N|\phi>\ =\ \phi(q\psi)$ with the prefactor in (38) unaltered and so,
\begin{eqnarray}
\psi <{\psi}^{\dagger}|F|\phi>&=& \frac{\phi(\psi)-\phi(-q\psi)}{1+q}.
\end{eqnarray}
This suggests to introduce q-differentiation as 
\begin{eqnarray}
\frac{d_q}{d_q\psi}\phi(\psi)&\equiv & \frac{\phi(\psi)-\phi(-q\psi)}{\psi (1+q)}.
\end{eqnarray}
Thus {\it{in the space of $\phi(\psi)$,  $F$ is represented by q-differentiation with respect to $\psi$}}. It can
be verified $\frac{d_q}{d_q\psi}{\psi}^n\ =\ [n]_f {\psi}^{n-1}$. In this way, we realize that the Bargmann spaces
for q-boson and q-fermion are very different. The monomials of quasi-Grassman variables
cannot be obtained by naive replacement of $q$ by $-q$. Now using these results, the expression (9) can be
written as 
\begin{eqnarray}
{\Big( \psi \frac{d_q}{d_q\psi}\Big)}^r\phi(\psi)&=&\sum_{s=1}^r {\cal{F}}^r_s (\psi)^s\Big( \frac{d_q}{d_q
\psi}\Big)^s\ \phi(\psi),
\end{eqnarray}
and the expression (24) gives
\begin{eqnarray}
{\Big( \frac{d_q}{d_q\psi}\ \psi \Big)}^r\ \phi(\psi)&=&\sum_{s=1}^r {\cal{B}}^r_s 
\Big( \frac{d_q}{d_q\psi}\Big)^s {\psi}^s
\ \phi(\psi).            
\end{eqnarray}  
Expressions (44) and (45) reveal the role of q-fermion Stirling number of the second kind and 'number
${\cal{B}}^r_s$' appearing in the anti-normal ordering of q-fermion operators, in expressing powers of
q-differential operators. 

\vspace{0.5cm}
    
{\noindent{\bf{7.Summary}}}

\vspace{0.5cm}

We have considered the q-fermion numbers introduced in the q-fermion oscillator algebra by Parthasarathy and
Viswanathan [6] in detail and obtained q-fermionic Stirling numbers of first and second kind explicitly. 
The q-fermionic Bell
number is obtained by means of q-fermionic Dobinsky formula. 
 q-fermionic Lah numbers are also considered. These results agree with those of Schork [4,5]. The case of
'anti-normal ordering' of q-fermionic annihilation and creation operators is studied and expansion coefficients
${\cal{A}}^r_s$ for q-bosonic operators and ${\cal{F}}^r_s$ for q-fermionic operators are introduced. Recurrence
relations for these are derived and these are very different from those encountered in q-stirling numbers or q-Lah
numbers. In this sense they are new. By taking matrix elements of the defining relations for these coefficients
between q-bosonic and q-fermionic Fock space states, we obtain expressions for the powers of q-bosonic and
q-fermionic numbers in terms of 'raising factorials'. These compliment the role of the q-Stirling numbers of the
second kind as they express powers of q-bosonic and q-fermionic numbers in terms of 'falling factorials'. The
theory of product densities of Ramakrishnan [10] is extended to q-stochastic point process and the necessity of
introducing q-product densities is emphasized. This leads to the identification of the effect of the degeneracy
with q-fermionic Stirling numbers of the second kind. We have given a Bargmann space representation of q-fermion
operators $F$ and $F^{\dagger}$ using q-fermion coherent states. This representation is used to express powers of  
 q-differential operators acting on the space of entire functions of quasi-Grassmann variable
as a series.  

{\vspace{0.5cm}}

{\noindent{\bf{References}}}

\vspace{0.5cm}

\begin{enumerate}
\item L.Carlitz, Trans.Amer.Math.Soc., {\bf{35}} (1933) 122; H.W.Gould, Duke. Math.J., {\bf{28}} (1961) 281;   
S.C.Milne, Trans.Amer.Math.Soc. {\bf{245}} (1978) 89.
\item J.Katriel and M.Kibler, J.Phys.A:Math.Gen. {\bf{25}} (1992) 2683.
\item A.J.Macfarlane, J.Phys.A:Math.Gen. {\bf{22}} (1989) 4581. \\
      L.C.Biedenharn, J.Phys.A:Math.Gen. {\bf{22}} (1989) L873.
\item M.Schork, J.Phys.A:Math.Gen. {\bf{36}} (2003) 4651.
\item M.Schork, J.Phys.A:Math.Gen. {\bf{36}} (2003) 10391.
\item R.Parthasarathy and K.S.Viswanathan, J.Phys.A:Math.Gen. {\bf{24}} (1991) 613.
\item J.Katriel, Phys.Lett. {\bf{A273}} (2000) 159.
\item R.Parthasarathy and R.Sridhar, in {\it{Stochastic Point Processes}}, Editors: S.K.Srinivasan and
A.Vijayakumar, Narosa Publishing House (New Delhi), 2003. 
\item G.E.Andrews, {\it{The Theory of Partitions}}, (London: Addison-Wesley), 1976. \\
      L.Comtet, {\it{Advanced Combinatorics}}, (Dordrecht: Reidel), 1974.
\item A.Ramakrishnan, Proc.Cambridge. Phil.Soc. {\bf{46}} (1950) 595; {\it{ibid}} {\bf{48}} (1952) 451;
{\it{ibid}} {\bf{49}} (1953) 473.
\item P.Narayana Swamy, {\it{Preprint}}, (1999), quant-ph/9909015.
\item M.Daoud, Y.Hassouni and M.Kibler, in: B.Gruber, M.Ramek (Eds). {\it{Symmetries in
Science X}}, Plenum, New York, 1998; M.Daoud and M.Kibler, Phys.Lett. {\bf{A321}}
(2004) 147.  
\item K.S.Viswanathan, R.Parthasarathy and R.Jagannathan, J.Phys.A:Math.Gen. {\bf{24}} (1992) L335. 
\item C.G.Wagner, {\it{Preprint}} http://www.math.utk.edu/~wagner/papers/paper4.pdf 
\item P.Blasiak, K.A.Penson and A.Solomon, Phys.Lett. {\bf{A309}} (2003) 198.  
\item A.J.Bracken, D.S.MacAnally, R.B.Zhang and M.D.Gould, J.Phys.A:Math.Gen. {\bf{24}} (1991) 1379.
\item R.Parthasarathy, {\it{Preprint}} IMSc.93/23, April 6, 1993.  
\end{enumerate}            
\end{document}